\documentclass[reprint,aip,jcp,amsmath,amssymb]{revtex4-1}
\pdfoutput=1
\usepackage[utf8]{inputenc}
\usepackage{amsmath}
\usepackage{graphicx}   
\usepackage{dcolumn}    
\usepackage{bm}         
\usepackage{textcomp}
\usepackage{amssymb}
\usepackage{array,multirow}
\usepackage{microtype}
\usepackage{mathtools}

\begin{document}
\bibliographystyle{tim}
\newcommand{\cev}[1]{\reflectbox{\ensuremath{\vec{\reflectbox{\ensuremath{#1}}}}}}

\title{Competing quantum effects in the free energy profiles and diffusion rates of hydrogen and deuterium molecules through clathrate hydrates}

\author{Joseph R. Cendagorta}
\author{Anna Powers}
\affiliation{Department of Chemistry, New York University, New York, NY, 10003, USA}
\author{Timothy J. H. Hele}
\affiliation{Department of Chemistry and Chemical Biology, Cornell University, Ithaca, New York 14853, USA. On intermission from Jesus College, University of Cambridge, CB5 8BL, UK.}
\author{Ondrej Marsalek}
\affiliation{Department of Chemistry, 333 Campus Drive, Stanford University, Stanford, CA 94305, USA.}
\author{Zlatko Ba\v{c}i\'{c}}
\thanks{zlatko.bacic@nyu.edu}
\affiliation{Department of Chemistry, New York University, New York, NY, 10003, USA}
\affiliation{NYU-ECNU Center for Computational Chemistry at NYU Shanghai, 3663 Zhongshan Road North, Shanghai, 200062, China}
\author{Mark E. Tuckerman}
\thanks{mark.tuckerman@nyu.edu}
\affiliation{Department of Chemistry, New York University, New York, NY, 10003, USA}
\affiliation{Department of Chemistry and Courant Institute of Mathematical Sciences, New York University, New York, NY 10003, USA.}

\date{\today}

\begin{abstract}
Clathrate hydrates hold considerable promise as safe and economical materials for hydrogen storage. Here we present a quantum mechanical study of H$_2$ and D$_2$ diffusion through a hexagonal face shared by two large cages of clathrate hydrates over a wide range of temperatures. Path integral molecular dynamics simulations are used to compute the free-energy profiles for the diffusion of H$_2$ and D$_2$ as a function of temperature. Ring polymer molecular dynamics rate theory, incorporating both exact quantum statistics and approximate quantum dynamical effects, is utilized in the calculations of the H$_2$ and D$_2$ diffusion rates in a broad temperature interval. We find that the shape of the quantum free-energy profiles and their height relative to the classical free energy barriers at a given temperature, as well as the rate of diffusion, are profoundly affected by competing quantum effects: above 25 K, zero-point energy (ZPE) perpendicular to the reaction path for diffusion between cavities decreases the quantum rate compared to the classical rate, whereas at lower temperatures tunneling outcompetes the ZPE and as result the quantum rate is greater than the classical rate.
\end{abstract}

\maketitle

\section{Introduction}
Clathrate hydrates are crystalline solids where guest molecules are encapsulated inside, and stabilize, the  close-packed polyhedral cavities within the three-dimensional (3D) host framework of hydrogen-bonded water molecules.\cite{MAO07B,MAO07A,SLOAN98} Nearly two decades ago it was demonstrated that, contrary to the long-held opinion, molecular hydrogen is capable of forming clathrate hydrates.\cite{DYADIN99,MAO02} Simple hydrogen hydrates, made of only H$_2$ and H$_2$O, adopt the classical structure II (sII).\cite{MAO07B,MAO07A} Its unit cell is cubic and contains 136 water molecules forming the hydrogen-bonded network.  This network is comprised of two types of cages: sixteen dodecahedral $(5^{12})$, or ``small'' cages, with 12 pentagonal faces, and eight hexakaidecahedral $(5^{12}6^4)$, or ``large'' cages, having 12 pentagonal and 4 hexagonal faces. Neutron diffraction studies of the pure D$_2$ clathrate hydrate\cite{MAO04A} found only one D$_2$ molecule in the small cages over the entire range of temperatures and pressures tested, and up to four D$_2$ molecules in the large cages, depending on the pressure. Hydrogen hydrates have attracted a great deal of interest as potential hydrogen storage materials\cite{MAO07B,MAO04,SCHUETH05,HU06,MAO07A,STROBEL09A} that would be environmentally friendly, safe, and economical. One of the main obstacles currently limiting the hydrogen storage applications of simple H$_2$ hydrates are the rather extreme conditions, high pressure and low temperature, required for their formation, as well as the low temperatures necessary to store them.  It has been shown that the addition of a second, larger promoter molecule such as tetrahydrofuran (THF) reduces drastically the formation pressure from 200 to 5 MPa at 280 K,\cite{SLOAN04,RIPMEESTER05} albeit at the cost of reduced hydrogen storage capacity. The binary THF + H$_2$ hydrate can also adopt the sII structure, where the THF molecules reside in the large cages, leaving only the small cages to be singly occupied by the H$_2$ molecules.

In addition to their promise as hydrogen storage media, hydrogen hydrates constitute an exceptional nanoscale ``laboratory'' for both theoretical and experimental investigations of the intriguing dynamical issues arising from the encapsulation of one or more light molecules in very confined spaces of rather high symmetry. The confinement results in the discretization of the translational, center of mass (c.m.) degrees of freedom of the guest molecules, and their coupling to the  already quantized rotational states. Both the discrete translational and rotational eigenstates are well separated in energy, due to the small mass and large rotational constant of the hydrogen molecule, as well as the nanoscale size of the confining spaces.  Consequently, the coupled translation-rotation (TR) energy level structure is sparse. It is even sparser due to the fact that the two protons of H$_2$ have the nuclear spin 1/2 and are fermions. The Pauli principle requires the total molecular wave function to be antisymmetric with respect to the nuclear exchange. This gives rise to two nuclear spin isomers, {\it para}-H$_2$ ({\it p}-H$_2$) and {\it ortho}-H$_2$ ({\it o}-H$_2$), having antisymmetric $I=0$ and the symmetric $I=1$ total nuclear spin states, respectively. For {\it p}-H$_2$, only even rotational quantum numbers $j$ are allowed, while {\it o}-H$_2$ can have exclusively odd rotational quantum numbers. As a result of these combined quantum effects, the dynamics of hydrogen molecules encapsulated in the clathrate cages is in the strongly quantum regime, {\it i.e}., it cannot be described accurately in terms of classical mechanics, particularly at the low temperatures at which the hydrogen hydrates are synthesized and at which the experiments on them are typically carried out. 

The salient features of the quantum TR dynamics of one or more hydrogen molecules encapsulated in the small and large cages of the sII clathrate hydrate were revealed in a series of rigorous theoretical studies over the past decade.\cite{BACIC06B,BACIC07C,BACIC08A,BACIC08C,BACIC09A,BACIC10B,BACIC16} The near-quantitative agreement between the computed TR energy levels and the early inelastic neutron scattering (INS) spectra recorded for binary THF + H$_2$ and THF + HD sII hydrates\cite{ULIVI07,ULIVI08} allowed the preliminary assignment of the latter. The synergy between theory and experiment, INS in particular, was strengthened significantly by the recent development of the quantum methodology for rigorous calculation of the INS spectra of a hydrogen molecule confined inside an arbitrarily shaped nanoscale cavity.\cite{BACIC11B,BACIC11C,BACIC13A} Quantum simulations using this methodology have proven capable of reproducing virtually every peak in the rich structure of the low-temperature INS spectra of the binary H$_2$\cite{BACIC13F,BACIC16} and HD sII hydrates,\cite{BACIC13A} leading to their complete analysis and unambiguous assignment.

The above INS experiments as well as the quantum calculations of the TR eigenstates and the INS spectra probe the dynamical behavior of H$_2$ confined, localized inside the cages of clathrate hydrates. However, there is ample evidence that the confinement is not complete, and that H$_2$ molecules can diffuse in both simple sII H$_2$ hydrates and binary sII THF+H$_2$ hydrates.  For the simple hydrogen hydrate at ambient pressure, the H$_2$ occupancy of the large cage was found to decrease gradually from the maximum of four to about two as the temperature increases from 80 to 160 K.\cite{MAO04A} Moreover, when the simple H$_2$ hydrate was subject to repeated cycles of heating to 150 K and quenching in liquid nitrogen, the average H$_2$ occupancy of the large cages diminished from four to three to two molecules.\cite{STROBEL09} In both experiments, the clathrate crystal structure remained intact, demonstrating that the diffusive migration of H$_2$ must take place,\cite{STROBEL09} and the small cages always remained singly occupied by H$_2$.\cite{MAO04A,STROBEL09} The $^1$H NMR measurements on simple H$_2$ hydrates have revealed the onset of H$_2$ diffusion above 120 K. \cite{CONRADI07,CONRADI08} The most plausible diffusive pathway for H$_2$ in the simple sII H$_2$ hydrate involves an extended network of hexagonal face sharing large cages, bypassing the small cages that have only pentagonal faces.\cite{STROBEL09} This is in qualitative accord with the calculated energy barriers for H$_2$ migration through pentagonal and hexagonal faces of 25--29 and 5--6 kcal/mol, respectively.\cite{RIPMEESTER07,KROES07} Other experimental studies, however, have suggested that the diffusion of H$_2$ in and out of the small hydrate cages is possible and can even be facile. When a pre-formed crystalline sII THF hydrate is exposed to hydrogen gas,\cite{RIPMEESTER07B, MULDER08} it absorbs H$_2$ rapidly at moderate pressures and at 250-265 K, resulting in the binary sII THF+H$_2$ hydrate. In the sII THF hydrate, all the large cages are singly occupied by the THF molecules, while the small cages are empty. Therefore H$_2$ can diffuse into the THF hydrate crystal only through the vacant  small cages, indicating that H$_2$ molecules must pass through the pentagonal faces shared by neighboring small cages. The activation energy measured for H$_2$ diffusion, about 0.7 kcal/mol,\cite{RIPMEESTER07B} is much smaller than the energy barrier of 25-291 kcal/mol calculated for the migration of H$_2$ through a pentagonal face.\cite{RIPMEESTER07}

Low-temperature diffusion of H$_2$ molecules in a complex chemical environment with a complicated structure and elaborate tunneling pathways is clearly a challenging problem of both fundamental and practical significance. Classical molecular dynamics (MD) simulations have been performed to investigate the free energies and diffusion of H$_2$ in the simple sII H$_2$ hydrate\cite{KROES07,MACELROY12,MACELROY13,Trin2015} and also in the binary sII THF+H$_2$ hydrate.\cite{MACELROY12,MACELROY13,Trin2015,ENGLISH16} In two of these studies,\cite{Trin2015, ENGLISH16} the free-energy barriers were computed for different H$_2$ occupancy of the neighboring large cages, and the barrier heights were found to decrease rather strongly with increasing H$_2$ occupancy. 

The major shortcoming of the classical MD simulations is the lack of the explicit treatment of the quantum effects, which should be significant for both the diffusive dynamics of H$_2$, given the highly quantum nature of its TR dynamics, and the motions of the framework of H$_2$O molecules, particularly at low temperatures. Previously, quantum effects have been shown computationally to be important for H$_2$ diffusion below about $120$ K in zeolite Rho,\cite{BHATIA05,BHATIA06} microporous aluminophosphate AlPO$_4$-25,\cite{kum08} a carbon molecular sieve (also experimentally),\cite{BHATIA010} and in a MOF material at 77 K.\cite{JOHNSON08} In the MD simulations,\cite{BHATIA05,BHATIA06,JOHNSON08,kum08} quantum effects were incorporated by means of the Feynman-Hibbs effective potentials\cite{FEYNMAN65} and transition state theory\cite{eyr35,eyr35rev} was also employed.\cite{BHATIA010} The first attempt to estimate the quantum tunneling contribution to H$_2$ migration between the cages of clathrate hydrate was based on the consideration of the 1D Eckart barrier.\cite{RIPMEESTER07} Very recently, Burnham and English,\cite{ENGLISH16} in addition to classical MD simulations, used path integral molecular dynamics (PIMD) to compute the free-energy profile for the diffusion of H$_2$ between two neighboring large cages of clathrate hydrate at the single temperature of 200 K, and obtained the quantum free-energy barrier that was $\sim$0.5 kcal/mol higher than the classical. This difference between the quantum and classical barrier heights is chemically significant at low temperatures.

In this article, we undertake a comprehensive study of the quantum statistical and dynamical effects in the diffusion of a single H$_2$ and D$_2$ molecule through the hexagonal faces of the adjacent large cages of clathrate hydrate over a wide range of temperatures. The free-energy profiles for H$_2$ and D$_2$ diffusion are generated by means of the PIMD simulations for temperatures ranging from 8 K to 200 K. These calculations reveal that the shapes of the quantum free-energy profiles and their heights relative to the classical profiles vary greatly with temperature, owing to the subtle interplay between two competing quantum effects: tunneling and the zero-point energy of the motions perpendicular to the reaction coordinate. The balance between the two shifts strongly as a function of temperature. The rates of H$_2$ and D$_2$ diffusion in the same temperature interval are calculated  using ring polymer molecular dynamics rate theory\cite{cra04a,cra05a,cra05b} which incorporates exact quantum statistics and approximate quantum dynamics.\cite{hel15a,hel15b} The computed rate is independent of the location of the dividing surface\cite{cra05b} and does not require presumption of one-dimensional dynamics.

The article is structured as follows. In section~\ref{sec:theory} we summarize RPMD rate theory and computational details. Results are presented in section~\ref{sec:results} and conclusions in section~\ref{sec:conclusions}.

\section{Theory}
\label{sec:theory}
\subsection{RPMD rate theory}
Since the energy barrier to hopping is significant compared to the thermal energy, diffusion is a rare event process and amenable to computation using rate theory.\cite{sul12a,mar08b} Reaction rates are calculated using Ring Polymer Molecular Dynamics (RPMD) rate theory,\cite{cra04a,cra05a,cra05b} which has been extensively reviewed elsewhere;\cite{hab13} here, we sketch the details relevant to this application.

We use the Bennett-Chandler factorization\cite{fre02} where the RPMD rate is the product of a static calculation and a time-dependent transmission coefficient,
\begin{equation}
k_{\rm RPMD}(T) = \lim_{t\to\infty} k_{\rm QTST}(T) \kappa(t).
\label{eq:rate}
\end{equation}
The static quantum transition-state theory rate $k_{\rm QTST}(T)$ is the instantaneous thermal flux through the dividing surface\cite{alt13a,hel13a,hel13b} 
\begin{equation}
k_{\rm QTST}(T) = \frac{1}{\sqrt{2\pi\beta \mu}}p(q_c^{\ddagger}),
\label{eq:qtst}
\end{equation}
where $\beta \equiv 1/k_{\rm B}T$ is the inverse temperature, $\mu$ is the reduced mass of the reaction co-ordinate, $\mu \equiv \left(\sum{\frac{1}{m_i}\mathbf{\left|\frac{\partial q}{\partial r_i}\right|^2}}\right)^{-1}$, and $p(q_c^{\ddagger})$ is the probability per unit length of finding the ring polymer reaction co-ordinate $q_c$ at the dividing surface $q_c^\ddag$, which is related to the corresponding free energy $A(q_c^{\ddag})$ by the relation 
\begin{equation}
p(q_c^{\ddagger}) = \frac{e^{-{\beta}A(q_c^\ddag)}}{\int_{q_0}^{q^{\ddag}}{e^{-{\beta}A(q_c)}}dq_c}.
\end{equation}
Here, the integral in the denominator is over the entire reactant region. The subscript $c$ denotes that the centroid [see Eq.~\eqref{eq:cen}] is used to form the reaction co-ordinate, in which case $k_{\rm QTST}(T)$ is formally equivalent to centroid-TST.\cite{gil87,gil87hyd,vot89rig}

The transmission coefficient $\kappa(t)$ accounts for recrossing of the dividing surface by RPMD trajectories, 
\begin{equation}
\kappa(t) = \frac{\left<\delta(q_c-q_c^{\ddagger})(p_c/\mu)h[q_c(t)-q_c^{\ddagger}]\right>}{\left<\delta(q_c-q_c^{\ddagger})(p_c/\mu)h(p_c)\right>}
\end{equation}
where $\langle \ldots \rangle$ is the thermal ring polymer expectation value, $\delta(q_c-q_c^{\ddagger})(p_c/\mu)$ is the ring polymer flux through $q_c^\ddag$ at $t=0$, and the heaviside function $h[q_c(t)-q_c^{\ddagger}]$ returns zero if the ring polymer centroid is in the reactant region at time $t$ and one if in the product region.

The overall RPMD rate is independent of the location of the dividing surface,\cite{cra05b} although it is computationally favourable to choose a `good' dividing surface, {\it i.e.,} one that minimizes recrossings and maximises $\kappa(t)$.\cite{sul11} Conveniently, the classical rate can be calculated using the above methodology but with a single path-integral bead.

\subsection{Computational details}
The system of interest contained 136 water molecules following the crystal structure of Mak et al.\cite{mak65}  The unit cell is a 1x1x1 with a box length of 17.31 {\AA}. Only one large cage is occupied by a single H$_2$ or D$_2$ hydrogen molecule. All other cages, large and small, are empty. In order to determine the free energy profile for the transfer of the hydrogen molecule between two neighboring large cages through a shared hexagonal face, the centroid-defined reaction co-ordinate, $q_c$, is used,
\begin{equation}
q_c(\mathbf{r_c}) = \left[\frac{1}{2}(\mathbf{r}^{\mathrm{H_a}}_c + \mathbf{r}^{\mathrm{H_b}}_c) - \mathbf{R}_{\mathrm{A}}\right] \cdot {\bm \mu}_{\mathrm{AB}} - \frac{|\mathbf{R}_{\mathrm{B}} - \mathbf{R}_{\mathrm{A}}|}{2},
\label{eq:qc}
\end{equation}
such that atomic positions are the centroids of their respective thermal paths; 
\begin{equation}
\mathbf{r}_c = \frac{1}{P}\sum_{k=1}^P \mathbf{r}_k \label{eq:cen}
\end{equation}
where $P$ is the number of path-integral beads.
This reaction coordinate measures the projection of the hydrogen molecule's c.m. vector, $(\mathbf{r}^{\mathrm{H_a}} +  \mathbf{r}^{\mathrm{H_b}})/2$ onto the axis generated by the centers of the two cages, denoted ($\mathbf{R}_{\mathrm{A}}$ and $\mathbf{R}_{\mathrm{B}}$), respectively.  Here, ${\bm \mu}_{{\mathrm{AB}}} = (\mathbf{R}_{\mathrm{B}}-\mathbf{R}_{\mathrm{A}})/|\mathbf{R}_{\mathrm{B}}-\mathbf{R}_{\mathrm{A}}|$.
In order to ensure that the collective variable is zero at the transition state, half the distance between the two cages is subtracted from this projection. 

Path integral molecular dynamics (PIMD) simulations~\cite{CHEM_PIMD_Tuckerman1} were run to generate the free-energy profiles, transforming the primitive bead coordinates to normal mode coordinates. Specifically, the blue moon ensemble\cite{car89,spr98,lar94} with a centroid defined collective  variable were used to calculate the free energy profiles, which entails running constrained MD simulations at numerous data points along the reaction coordinate within the canonical ensemble (NVT).   Each constrained simulation was equilibrated for 100 ps, followed by a 400 ps production run with a time step of 0.25 fs. All simulations used 48 path-integral beads, which was found to be sufficient to converge the statistics down to 8 K. 

As in previous studies,\cite{BACIC08C,BACIC10B} the interaction between the water molecules of the clathrate cages and the single hydrogen molecules were assumed to be pairwise additive. However, unlike previous studies the q-TIP4P/F\cite{hab09} water model was used to describe the interactions of the water molecules. This model is a modified version of TIP4P/2005,\cite{aba05} which implicitly accounts for quantum effects. By contrast, PIMD simulations must be used to account for quantum effects in the q-TIP4P/F model, which is why it has been chosen for the present study.  The H$_2$-H$_2$O interaction potential consists of a sum of Coulombic and Lennard-Jones contributions. The point charges for the water molecules were derived from the q-TIP4P/F model, which is a 4-site, 3-point charge model. The hydrogen molecule charges are those from Alavi et al,\cite{ala05a} which include a positive charge of +0.4238 $e$ on the hydrogen atoms and a negative charge of -0.9864 $e$ on the c.m. of the H$_2$ molecule. The Lennard-Jones potential acts solely between the c.m. of the hydrogen molecule and the oxygen atoms of the water molecules. These parameters are determined following the standard Lorentz-Berthelot combination rules. In the original model of H$_2$ from Alavi et al.,\ the molecule was treated as rigid, which complicates the path integral simulations. Thus, as was done previously,\cite{BACIC10B} a harmonic bond was introduced between the hydrogen atoms of the H$_2$ molecule with a spring constant of $4.177 \times 10^5$ K/\AA$^2$ and equilibrium bond length of 0.74 \AA.
Nose-Hoover chains\cite{mar92} were used to globally thermostat the centroid mode of the ring polymer and massively thermostat the remaining modes, meaning that a separate thermostat chain is attached to each Cartesian direction of each of the remaining modes. The long-range electrostatic interactions were calculated using the smooth Particle-Mesh Ewald method.\cite{dar93} The free energy profiles were calculated by taking data from snapshots every 5 fs. In the classical molecular dynamics (MD) simulations, the number of ring polymer beads was reduced to one. All simulations were performed using the PINY\_MD molecular dynamics package.\cite{tuc00}

For the transmission coefficient ``parent'' simulations, the protocol was the same as the one above for calculating the free energy, except the production runs were only 250 ps and the reaction coordinate was constrained to the dividing surface ($q_c = 0$). Twenty different parent simulations were performed for each of the seven temperatures and a child trajectory was chosen randomly every 1-2 ps. For the ``child'' simulations, Ring Polymer Molecular Dynamics (RPMD)\cite{cra04a} was performed (NVE simulations) in the normal mode coordinates, which are similar to the NVT simulations, except all thermostats are turned off, and the mass assigned to each bead is the physical mass of the atom. For higher temperatures ($\geq$ 25 K), the RPMD simulations were run for 1 ps, while the lower temperatures were run for much longer times, specifically 5 ps, 7.5 ps and 15 ps for 17 K, 12 K, and 8 K respectively, in order for the transmission coefficient to reach a plateau. For the D$_2$ RPMD simulations, temperatures $\geq$ 25 K were run for 1 ps while the 8 K simulations were run for 7.5 ps. Each parent simulation produced at least 160 children trajectories and thus a total of 3200 configurations were used for each temperature. This constituted a sufficient number of configurations to converge the transmission coefficient at all temperatures.

\section{Results}
\label{sec:results}
We first explore the free energy profiles generated from PIMD simulations from which the QTST rate can be calculated using Eq.~\eqref{eq:qtst}. We then present transmission coefficient calculations before discussing the quantum and classical diffusion rates.

\begin{figure*}[tb!]
 \centering
 \includegraphics[width=\textwidth]{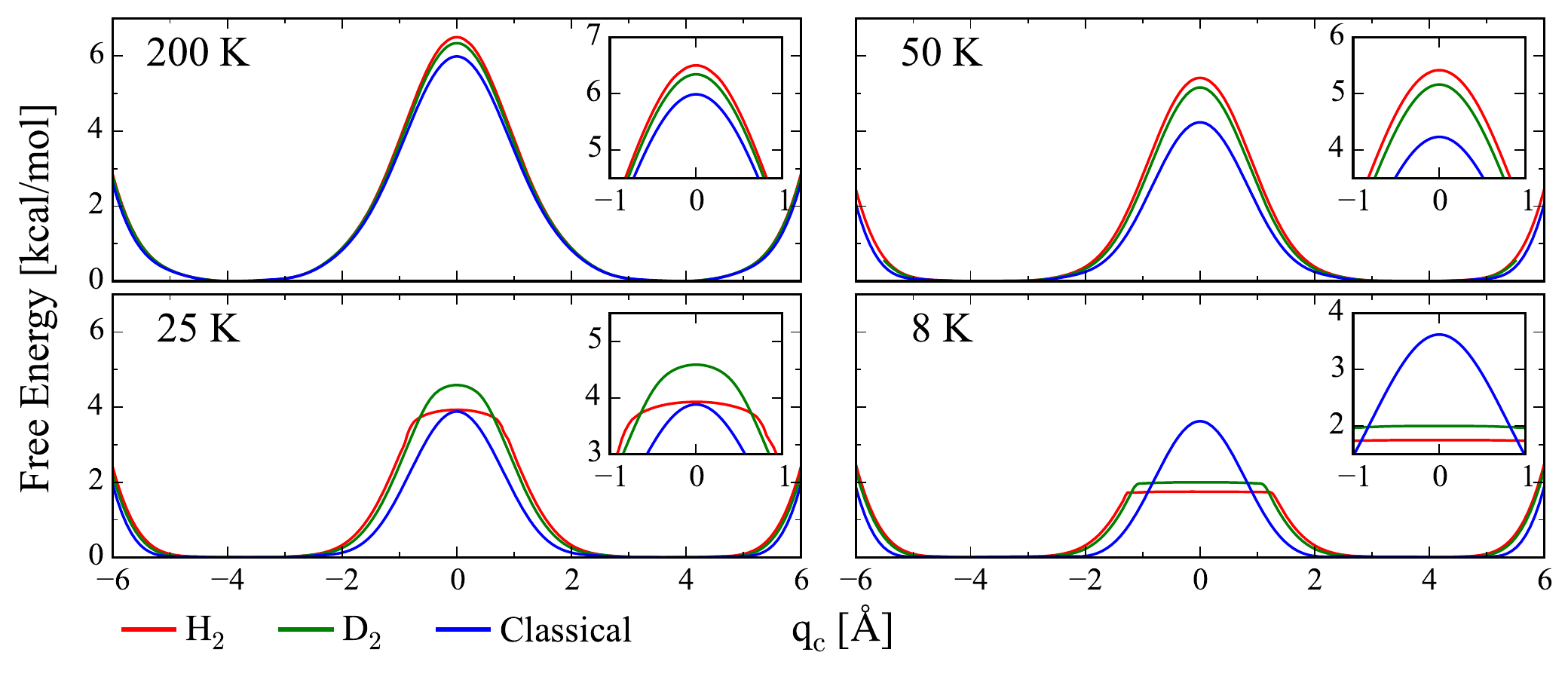}
 \caption{Free energy profiles for a single hydrogen/deuterium molecule moving through a hexagonal face between two large cages. The H$_2$ (red) and D$_2$  (green) results are from PIMD calculations, which when performed with a single path-integral bead give the classical (blue) free energies. The insets in each plot provide a clearer depiction at the dividing surface.}
 \label{fgr:FreeEnergy}
\end{figure*}

Figure~\ref{fgr:FreeEnergy} shows the quantum free-energy profiles for the diffusion of a single H$_2$ or D$_2$ molecule from one large cage to another, along with the results of classical calculations for H$_2$ in the temperature range 8--200 K. In addition to the temperatures shown, both quantum and classical H$_2$ simulations were undertaken at additional temperatures of 100K, the results of which were found to lie between 200 K and 50 K, and 17 K and 12 K, the results of which were found to lie between those for 25 K and 8 K (see Supplementary Information). Note that no quantum simulations for D$_2$ were performed at these three additional temperatures. Between 50 K and 200 K, the free energy profiles are qualitatively similar, with the barrier heights being ordered as H$_2> $D$_2 >$ classical. This suggests that there is considerable zero-point energy (ZPE) perpendicular to the reaction coordinate and that the swelling of the H$_2$ (or D$_2$) ring polymer with decreasing temperature constricts its passage through the hexagonal `bottleneck' between the clathrate cages. We also note that the height of the classical barrier decreases as the temperature is lowered (though not nearly as much as for H$_2$ or D$_2$) and attribute this to reduced thermal fluctuations of the clathrate water molecules' increasing the effective size of the hexagonal bottleneck.

The PIMD simulations by Burnham and English\cite{ENGLISH16} of the H$_2$ diffusion between two large clathrate hydrate cages at 200 K also yielded the quantum free-energy barrier higher than the classical one, by about 0.5 kcal/mol. This is in very good agreement with the difference between the quantum and classical free-energy barrier heights at 200 K computed in this study, which is approximately 0.51 kcal/mol. In addition, the larger quantum free-energy barrier height relative to the classical one and the decreasing difference between the two with increasing temperature were observed in simulations using the Feynman-Hibbs approach of the diffusion of H$_2$ (and D$_2$) through the narrow quasi-1D channels of AlPO$_4$-25.\cite{kum08} This supports the view that the ZPE contribution to the free-energy barrier arises from ``squeezing'', or localizing, the light H$_2$  molecule as it passes through the narrow pores, or openings, in the host material.

The free-energy profiles in Fig.~\ref{fgr:FreeEnergy} for temperatures between 8 K and 25 K differ qualitatively from those at higher temperatures. First, the quantum free-energy profiles begin to flatten; at 25 K the flattening for H$_2$ is slight and for D$_2$ is unnoticeable, but this is very pronounced for both H$_2$ and D$_2$ at 8 K. In contrast, the classical free-energy profiles change little in this temperature range. Second, at 25 K the classical and H$_2$ barrier are virtually identical in height, but the H$_2$ and D$_2$ barriers decrease rapidly with decreasing temperature, and at 8 K, they are approximately half the classical value. These phenomena imply that, around 25 K, quantum tunneling becomes the dominant mechanism for the barrier crossing, and its importance increases with decreasing temperature. 

\begin{figure*}[tb]
 \centering
 \includegraphics[width=\textwidth]{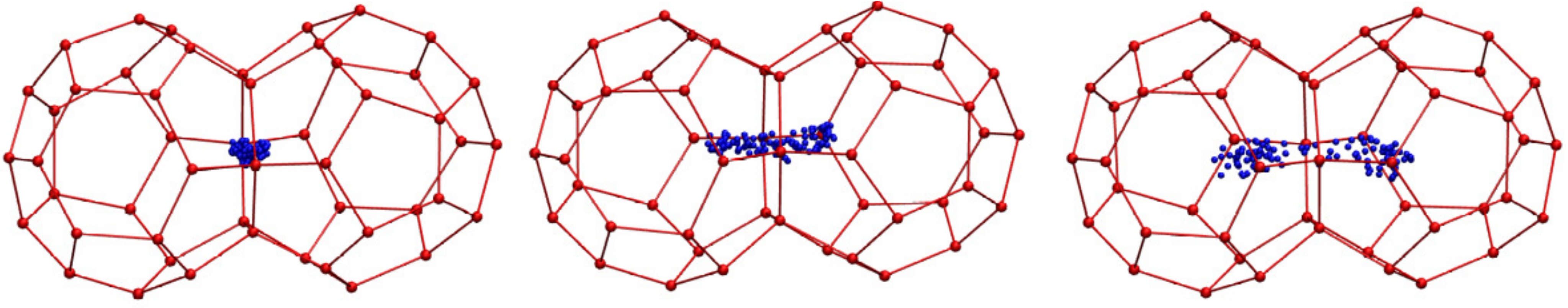}
 \caption{Snapshots of PIMD simulations showing the H$_2$ ring polymer (blue beads) at the dividing surface at 50 K, 25 K, and 8 K (left to right). There are a total of 96 beads corresponding to both H$_2$ atoms (48 beads per atom). The oxygen atoms are shown in red, forming the clathrate cages, and their associated hydrogen atoms are removed for clarity.}
 \label{fgr:Delocal}
\end{figure*}

The onset of deep tunnelling can be inferred from the crossover temperature\cite{ric09a} of 42 K for H$_2$ diffusion and 30 K for D$_2$ diffusion. Beneath this temperature, defined as
\begin{equation}
T_c = \frac{\hbar \omega_b}{2\pi k_{\rm B}}
\end{equation}
where $\omega_b$ is the (imaginary) barrier frequency, the saddle point on the ring polymer potential energy surface acquires one unstable mode. In qualitative terms, beneath $T_c$ the ring polymer with its centroid constrained to the dividing surface stretches significantly into the reactant and product region\cite{ric09a} as is seen in Fig.~\ref{fgr:Delocal} and leads to the observed flattening of the free energy profiles.

In order to illustrate the competing effects of ZPE and tunnelling, we show snapshots of PIMD simulations with the H$_2$ molecule constrained to the dividing surface in Fig.~\ref{fgr:Delocal}. At 50 K, the swelling of the ring polymer is confined to the transition state (the hexagonal bottleneck between the two clathrate cages), and therefore increases the free energy barrier. At 25 K some delocalization is noticeable whereas at 8 K the ring polymer is highly delocalised, with most of the beads near the center of one cavity or the other, and only a few near the bottleneck. This considerably decreases the free energy compared to a single bead (the classical result), meaning that tunnelling outcompetes ZPE. Furthermore, small movements of the centroid of the H$_2$ molecule do not appreciably alter the free energy since the majority of beads will still be delocalized in one cavity or the other, leading to a flattening of the free energy profiles as seen in Fig.~\ref{fgr:FreeEnergy}.

\begin{figure*}[tb]
\centering
  \includegraphics[width=\textwidth]{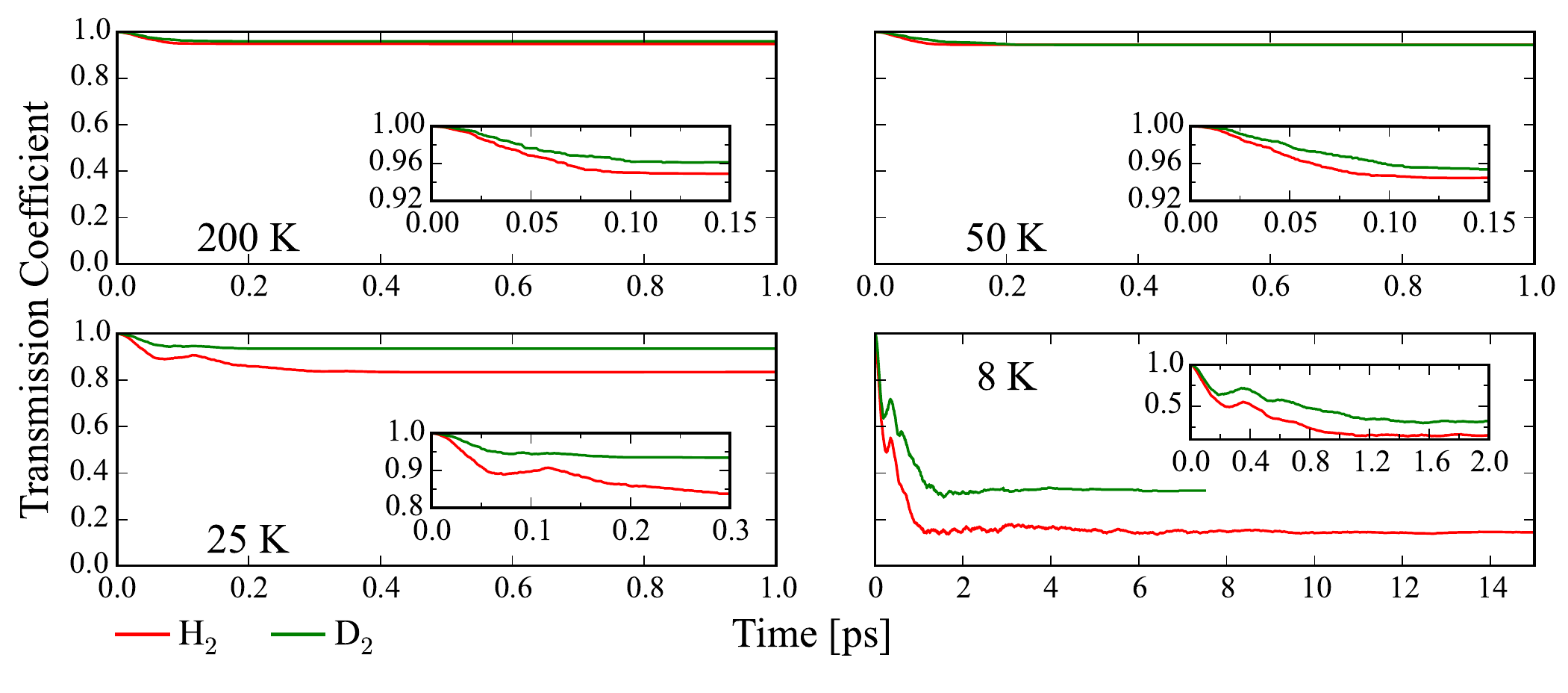}
  \caption{RPMD transmission coefficients for H$_2$ (red) and D$_2$ (green) over a range of temperatures. The $y$-axis is the same scale for all four graphs, but the $x$-axis is different at 8 K due to a longer plateau time; D$_2$ simulations were run for 7.5ps and H$_2$ simulations 15ps.}
  \label{fgr:Trans}
\end{figure*}

The transmission coefficients calculated from the RPMD simulations are shown in Fig.~\ref{fgr:Trans}. The classical transmission coefficients (1 bead, see SI) are near unity indicating very little recrossing dynamics. Furthermore, the transmission coefficients for the temperatures greater than 25 K are also near unity and converge rapidly.  Below 25 K the H$_2$ and D$_2$ transmission coefficients decrease markedly, indicating considerable recrossing of the ring polymer dividing surface. For a symmetric system like those considered here, the centroid is expected to be a `good' dividing surface (a good approximation to the optimal dividing surface that maximizes $\kappa(t)$) down to half the crossover temperature, which is 21 K for H$_2$ and 15 K for D$_2$. Beneath this temperature, the optimal dividing surface will become a function of the second ring polymer normal modes (the pair of normal modes of a free ring polymer\cite{cer10a} with frequency $\omega_{\pm 2} = 4\pi/\beta\hbar$ in the large $P$ limit).\cite{ric09a} Consequently, the centroid will be a suboptimal dividing surface and experience significant recrossing, as seen in Fig.~\ref{fgr:Trans}. We stress that although the dividing surface is no longer optimal and the QTST rate computed with it too high, the transmission coefficient accounts for this deficiency and the overall RPMD rate is independent of the dividing surface chosen.

Temperatures below 25 K require a much longer simulation time, as seen in Fig.~\ref{fgr:Trans}. All the transmission coefficients for temperatures lower than 25 K contain a shoulder, which varies with temperature and can be attributed to a ``tug of war'' of the ring polymer at short times prior to the plateau region.\cite{mar08}  
Due to the delocalization of the ring polymer (Fig.~\ref{fgr:Delocal}), some of the ring polymer is located in one cage while the rest is located in the other cage. Therefore, there is an initial tug of war in which the ring polymer moves between cages. The duration of this ``tug of war'' increases with decreasing temperature due to a slackening of the ring polymer springs. The $j$th normal mode of the ring polymer has a frequency (in the absence of an external potential) of $\omega_j = 2\pi j k_{\rm B} T/\hbar$, and therefore, the time at which the shoulder appears is expected to increase linearly with $1/T$, as is observed. However, eventually the ring polymer must end up in one cage or the other, leading to an eventual plateau of the transmission coefficient.  A much more pronounced tug of war was observed in the diffusion of hydrogen/muonium atom in ice;\cite{mar08b} repeated oscillations are probably washed out in the present case due to decoherence of the H$_2$ ring polymer dynamics, which here represent the motion of two hydrogen atoms rather than one.

\begin{figure}[tb]
\centering
  \includegraphics[width=8.1cm]{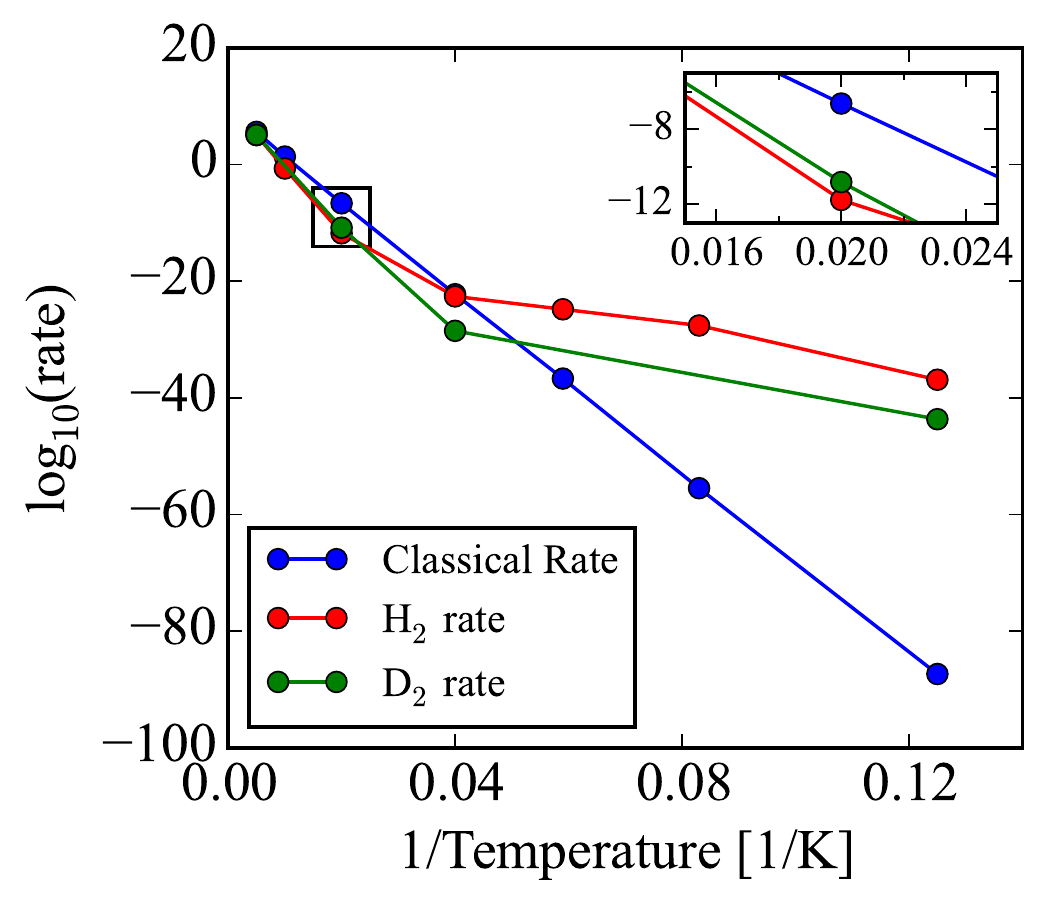}
  \caption{Arrhenius plot of the rate of H$_2$ diffusion from quantum RPMD (red), D$_2$ diffusion (green) and classical (blue) calculations. The inset focuses on 50 K, which demonstrates the inverse kinetic isotope effect.}
  \label{fgr:Arrhenius}
\end{figure}

Figure~\ref{fgr:Arrhenius} presents the H$_2$ and D$_2$ RPMD rates alongside the classical calculation for H$_2$ diffusion, all obtained by combining the free energy results and transmission coefficient calculations using Eq.~\eqref{eq:rate}. To within graphical accuracy, the quantum TST rates for D$_2$ and H$_2$ are identical to the RPMD rates, and the classical TST rates are identical to the classical rates, hence, they are not shown. This suggests that the main contribution to the temperature dependence of the rate is statistical (the free-energy barrier) rather than dynamical (recrossing) effects.

The standard linear dependence of the logarithm of the classical rate on inverse temperature is also shown in Fig.~\ref{fgr:Arrhenius} and is seen to lead to errors of $\sim 10^{50}$ at the lowest temperature 8 K. The quantum rates show two qualitatively different regions. For $T\ge 25$ K, the quantum rates are lower than the classical rate, attributable to the zero-point energy perpendicular to the reaction coordinate. Below 25 K, deep tunnelling reverses the situation, and the quantum rates become many orders of magnitude higher than the classical rates. Furthermore, at low temperatures, the D$_2$ rates are lower than those for H$_2$, exhibiting the conventional kinetic isotope effect. However, around 50 K (see inset) the D$_2$ rate is higher than the H$_2$ rate due the slightly lower free energy of D$_2$ at the dividing surface (see above), leading to an inverse KIE.

The rates calculated here for the diffusion of an H$_2$ molecule from a singly occupied large cage to a neighboring empty large cage cannot be directly compared with the experimentally measured diffusion rates. The free-energy barriers for H$_2$ diffusion have been found to differ greatly for different H$_2$ occupancy values of the large cages involved,\cite{Trin2015, ENGLISH16} implying that the diffusion rates, which have not yet been reported, will exhibit strong dependence on the cage occupancy as well. This means that in order to have a meaningful comparison with experiment, calculation of the diffusion rate of H$_2$ in bulk clathrate hydrate at a given temperature will require determining the quantum free-energy profiles for many possible combinations of H$_2$ occupancies of the adjacent large cages, using these to compute the rates of diffusion for each such combination, and finally averaging over these rates. Clearly, it is a computationally most demanding task.

\section{Conclusions}
\label{sec:conclusions}
A quantitative molecular-level understanding of the diffusion of hydrogen molecules in clathrate hydrates at low temperatures is essential for possible future applications of hydrates as hydrogen storage materials. In particular, it is important to account for quantum effects, which are significant in the case of light molecules diffusing through a highly structured nanoporous material.

In this article we have computed the quantum free-energy profiles and rates of diffusion of an H$_2$ and D$_2$ molecule through the hexagonal face of two neighboring large clathrate hydrate cavities in the temperature interval of 8 K to 200 K. PIMD simulations were employed to compute the free-energy profiles, while the diffusion rates were calculated using RPMD rate theory to incorporate quantum dynamical effects, and the classical rates were calculated on the same footing for comparison.

We find that the free-energy profiles for the diffusion of H$_2$ and D$_2$ are strongly affected by two competing quantum effects, tunneling and the ZPE associated with the motions of the diffusing hydrogen perpendicular to the reaction coordinate, and whose relative contributions vary greatly with temperature. At relatively high temperatures, but still in the non-classical regime, the dominant quantum effect is the ZPE of the transverse modes which enhances the free-energy barrier and hinders diffusion. Alternatively, this quantum effects can be viewed as the swelling of the ring polymer perpendicular to the reaction co-ordinate, increasing the effective size of the hydrogen molecule, and making it harder to diffuse through the bottleneck presented by the hexagonal face shared by the two large cages. At temperatures below 25 K, this effect is more than offset by deep tunneling.

The classical transmission coefficient calculations show virtually no temperature dependence, whereas the RPMD calculations at low temperatures illustrate a brief ``tug of war'' with the hydrogen ring polymer between the two clathrate cages leading to a much reduced transmission coefficient.

The reaction rates themselves exhibit a higher-temperature region, where quantum effects, primarily the ZPE of the transverse modes, decrease the H$_2$ and D$_2$ quantum rates relative to the classical rate, and an inverse KIE is observed, and a low-temperature region, where the quantum rates are substantially larger than those obtained from a classical calculation, owing to the contribution from tunneling, and a conventional KIE is seen. This suggests the possibility of quantum kinetic sieving either for H$_2$ or D$_2$ as a function of temperature.  

The work presented in this paper reveals that even the simplest diffusion process in hydrogen hydrates, the transfer of a hydrogen molecule from one large cage to the neighboring empty large cage through a shared hexagonal face, exhibits surprising complexity when quantum effects are taken into account and a wide temperature range is considered. This study represents an important and necessary first step, which demonstrates the capabilities of the quantum methodologies essential for the next, significantly more demanding stage of our investigations. In particular, the computation of realistic diffusion rates which can be compared to experimental results will require calculation of the quantum free-energy barriers and diffusion rates for a wide range of H$_2$ and D$_2$ occupancies of clathrate cages, and averaging over them. Computation of transmission coefficients may require advanced thermostatting techniques \cite{ros14a,hel15c,hel16a} in order to enhance statistical sampling. Diffusion through the pentagonal faces connecting small and large clathrate cages will also be investigated, as will be the effect of nuclear spin statistics\cite{miu00} on the diffusion rate at low temperatures. 

\section*{Acknowledgements}
AP acknowledges a Margaret and Strauss Kramer Fellowship. TJHH acknowledges a Research Fellowship from Jesus College, Cambridge. ZB and MET acknowledge partial support of this research by the National Science Foundation through the Grant CHE-1112292.

\bibliography{main,bacic,mark,tim}
\end{document}